\documentclass[preprintnumbers,amsmath,11pt,amssymb,floatfix,superscriptaddress,nofootinbib]{article}
\usepackage{array}
\usepackage{relsize}
\usepackage{cite}
\usepackage{amsmath}
\usepackage{amssymb}
\usepackage{graphicx}
\def\mytitle#1{\setcounter{equation}{0}
\setcounter{footnote}{0}
\begin{flushleft}\Large\textbf{#1}\end{flushleft}
\vspace{0.25cm}}
\def\myname#1{\leftline{{\large #1}}\vspace{-0.13cm}}
\def\myplace#1#2{\small\begin{flushleft}\textit{#1}\\
\texttt{#2}\end{flushleft}}

\def\myclassification#1{\small\noindent
PACS:
       #1\vspace{0.5cm}}
\usepackage{graphicx}
\begin{document}

\mytitle{Volumetric Maxima to be Attained by a Nonstatic Black Hole}

\myname{Sandip Dutta\footnote{duttasandip.mathematics@gmail.com}, Ritabrata Biswas\footnote{biswas.ritabrata@gmail.com} and Prasanta Choudhury\footnote{prasantachoudhury98@gmail.com}}
\myplace{Department of Mathematics, The University of Burdwan, Golapbag Academic Complex, City : Burdwan-713104, Dist. : Purba Barddhaman, State : West Bengal, India.}{} 
\begin{abstract}
Christodoulou and Rovelli have calculated maximal interior volume of a Schwarzschild black hole which linearly grows with time. Recently, the entropy of interior volume in a Schwarzschild black hole has also been calculated. In this article, the Eddington-Finkelstein metric is slightly modified. This modified metric satisfies Einstein's equations. The interior volume of a black hole is also calculated with the modified metric. The volume explicitly depends on a function of time, different from the Christodoulou and Rovelli volume. Also entropy is calculated corresponding to the volume which is proportional to the square of a function of time and thermodynamics is studied.\\ 

\end{abstract}

\myclassification{ 04.70 Dy, 04.70.-S, 02.40.Gh }


Black hole is a central singularity at $r=0$ and some singular horizon known as Cauchy horizon etc covered/wrapped by another singular horizon  which is called event horizon. Generally, if we consider Planck's unit, i.e., $G=c=\hbar=\kappa=1$, the radius of event horizon in our curved space time is twice of the mass of the central gravitating compact object. There arises an obvious question that how much volume of a black hole is occupied inside the event horizon. If we assume the problem in flat space, the volume inside a sphere with radius $r(=2m)$ is $\frac{4}{3}\pi(2m)^3$. But this intuition of flat space time becomes invalid for the curved geometry inside a black hole. The volume of the surface where the time coordinate $t$ is constant, depends on the arbitrary choice of coordinates. This is discussed by various authors in different references like \cite{Parikh1, Grumiller1, Ballik1, Ballik2, DiNunno1, Finch1, Cvetiˇc1, Gibbons1}. Recently, a different way of thinking about the volume inside a black hole is suggested by Christodoulou and Rovelli\cite{Christodoulou1} based on a simple observation that the exterior of Schwarzchild metric is static but the interior is not. This leads us to the notion that volume can be time dependent. The horizon is naturally decorated by two spheres. The definition of ``interior volume" is associated to the idea of single two sphere. Let us consider a two sphere ${\cal{S}}$ in Minkowski space time. The volume inside the two sphere ${\cal{S}}$ is equal to the volume of the largest spacelike spherically symmetric surface $\Sigma$ bounded by ${\cal{S}}$, where $\Sigma$ lies on the simultanity surface determined by ${\cal{S}}$. This characterisation of the volume inside the two sphere in flat space time also remains valid for the case of spherical black holes. By using this notion of volume inside a sphere Christodoulou and Rovelli\cite{Christodoulou1} have found a simple expression for volume(CR volume hereafter), given by
\begin{equation}
V(v)_{v \rightarrow \infty} = 3\sqrt{3} \pi m^2 v
\end{equation} 
by determining the maximal volume surface $\Sigma$ when $v$ is enough large respect to $m$.

After one gets the specific form of volume inside a black hole the requirement to know the physical significance of this volume is observed. Is it relevant to some information of black hole? It is inevitable to involve Hawking radiation\cite{Hawking1} and black hole thermodynamics\cite{Bekenstein1, Bardeen1} in such studies. Recently, Zhang\cite{Zhang1} made some quantitative calculations to estimate the entropy associated with the CR volume. If we consider a black hole with mass $m$, due to the emission of thermal radiation, the life time of the black hole becomes $\sim m^3$ and CR volume inside the black hole has an extra ordinary form $\sim m^5$. So one may arise the query on the number of field modes included in such a large volume when radiation happens and the background geometry is also changed according to the Einstein's equation, which is actually first law of thermodynamics.

In this letter,  First, we revisit some definitions and previous works about the volume inside the black hole and the entropy corresponding to the volume. Next,  we try to determine the volume from the modified Eddington-Finkelstein metric and also show some graphs of the maximal volume surface. Then we calculate entropy corresponding to the volume. Lastly we summarise the results.

We will consider the Eddington-Finkelstein coordinates $(v,r,\theta,\phi)$. To make it easy, we will choose a null spherical shell of energy $m$ collapsing towards the $v=0$ surface. Spacetime is found to be flat before this surface. After we cross this surface, the line element follows the standard Schwarzchild geometry. Schwarzchild black hole in Eddington-Finkelstein coordinates looks like
\begin{equation}\label{P4.1}
ds^2=-f(r,t)dv^2+2dvdr+r^2d\theta^2+r^2 sin^2\theta d\phi^2,
\end{equation}
where $f(r,t)=1-\frac{2ma(t)}{r}$ and $a(t)=(1+a_1t+a_2t^2+...+a_nt^n)$. This kind of metric was previously found by the reference \cite{M. Sharif_Wajiha Javed} where the black hole's mass was taken to be the function of $r$ and $t$ both. Black hole evaporation in a non commutating charged Vaidya metric has been studied. Here, we have to note some important information : $a_i$-s are parameters which are very small such that $a(t)$ may tend to $1$, whenever a terminal case is required. We consider the mass of the black hole is increasing with time as matters get accreted towards the black hole with time $t$. Also, universe's average density is considered to be very small. So, the black hole's mass increases gradually. Because of this, we assume that the $a_i$ parameters are very small, i.e., tends to zero. The partial derivative of $f(r,t)$ with respect to $t$ tends to zero, i.e., $\frac{\partial f}{\partial t}\approx 0$ if $a_i$-s are small. Easily, we can check that the Einstein's equation is satisfied. Also, the radial coordinates increment is very small. So, we consider that the mass of the BH is $ma(t)$ after a time $t$, i.e., mass of the BH is $m$ (at $t=0$). Also, we have assumed that $r$ varies from $2ma(t)$ to zero. The relation between the advanced time $v$ and the Schwarzchild coordinates $t,~r,~\theta,~\phi$ are written as $v=t+r+2ma(t) \log{\mid r-2ma(t)\mid}$. The benefit of the coordinate on the static surface is that there is no coordinate singularity at the event horizon. So, it can be logically continued to all $r>0$, which is appropriate for the description of the geometry of the collapsed matter.

$\Sigma$ is a $3D$ spherically symmetric surface which can be concerned as the direct product of a curve $\gamma$ and a 2-sphere in the $v-r$ plane, given by
\begin{equation}\label{P4.2}
\Sigma \equiv \gamma \times S^2
\end{equation}
\begin{equation}\label{P4.3}
\gamma \mapsto (v(\lambda), r(\lambda))
\end{equation} 
The curve $\gamma$ is taken as a parametric form, where $\lambda$ is an arbitrary parameter. On the horizon ($r=2m$), we assume $\lambda$ to vanish. $\lambda_f(f$ for `final') is the value of $\lambda$  at $r=0$. So, the primary and last end points of $\gamma$ are defined by
\begin{equation}\label{P4.4}
r(0)=2ma(t),~~~~~r(\lambda_f)=0
\end{equation} 
\begin{equation}\label{P4.5}
v(0)=v,~~~~~v(\lambda_f)=v_f
\end{equation}
$\lambda,~\theta,~\phi$ are the coordinates of the surface $\Sigma$. The metric on $\Sigma$ is defined by
\begin{equation}\label{P4.6}
ds_{\Sigma}^2=\left\{-f(r,t)\dot{v}^2+2\dot{v}\dot{r}\right\}d\lambda^2+r^2d\theta^2+r^2 sin^2\theta d\phi^2,
\end{equation}
where the differentiation with respect to $\lambda$ is indicated by the `dot'. For all coordinate values, we have to maintain that $ds^2{\mid_{\Sigma}}>0$ and we get the condition that $\Sigma$ is space like which looks like 
\begin{equation}\label{P4.7}
-f(r,t) \dot{v}^2+2 \dot{v} \dot{r}>0
\end{equation}
If $V_{\Sigma}[\gamma]$ is the proper volume of $\Sigma$, then it is defined as
\begin{equation}\label{P4.8}
V_{\Sigma}[\gamma]=\int_0^{\lambda_f}d\lambda \int_{S^2}d\Omega\sqrt{r^4(-f(r,t)\dot{v}^2+2\dot{v}\dot{r})sin^2\theta}=4\pi \int_0^{\lambda_f}d\lambda \sqrt{r^4(-f(r,t)\dot{v}^2+2\dot{v}\dot{r})}
\end{equation}
The volume is extremized by the surface $\Sigma_v$. Also, above integral is extremized by the curve. The volume is derived by the curve $\gamma_v$ and by the particularization of $v$ and $v_f$.

The Lagrangian equations of motion are
\begin{equation}\label{P4.9}
{\cal{L}}(r,v,\dot{r},\dot{v})=\sqrt{g_{\alpha \beta} dx^{\alpha} dx^{\beta}}=\sqrt{r^4(-f(r,t)\dot{v}^2+2\dot{v}\dot{r})}.
\end{equation} 
The concerned metric takes the form
\begin{equation}\label{P4.11}
ds^2_{M_{aux}}=r^4\left\{-f(r,t)dv^2+2dvdr\right\}.
\end{equation}
Finding the geodesics of the above metric is equivalent to find $\Sigma_v$.
Also, the equation (\ref{P4.8}) denotes the proper length of the geodesic in the above auxiliary metric (\ref{P4.11})(times $4\pi$) which is exactly the volume of $\Sigma$.

$\Sigma$ acts as space like from the condition (\ref{P4.7}). This indicates that ${\cal{L}}>0$, since $r\geq 0$. At $r=0$, the Lagrangian is identical to zero, which is the last point for the geodesic. In ${\cal{M}}_{aux}$, $\gamma$ is a space like geodesic. A well fitted parametrization is chosen as $\lambda$, as the proper length in ${\cal{M}}_{aux}$. After the maximization, we set
\begin{equation}\label{P4.12}
{\cal{L}}(r,v,\dot{r},\dot{v})=1\Rightarrow r^4(-f(r,t)\dot{v}^2+2\dot{v}\dot{r})=1
\end{equation}
and from (\ref{P4.8}) we instantly have
\begin{equation}\label{P4.13}
V=4\pi \lambda_f
\end{equation}
Also, $\xi^{\mu}=(\partial_v)^{\mu} \propto (1,0)$ is the Killing vector of the metric $g_{\alpha \beta}$. Again the inner product of $\xi$ and its tangent $\dot{x}^{\alpha}(\dot{v},\dot{r})$ are conserved. As $\gamma$ is affinal parametrized geodesic in $M_{aux}$, so we get
\begin{equation}\label{P4.14}
r^4(-f(r,t)\dot{v}^2+2\dot{v}\dot{r})=A.
\end{equation}
However, we can see from the equation (\ref{P4.14}) that $\dot{r}$ turns to be infinite.
Hence from the equation (\ref{P4.12}) and (\ref{P4.14}), we can recast it in the following form
\begin{equation}\label{P4.15}
\dot{r}=-r^{-4}\sqrt{A^2+r^4f(r,t)}
\end{equation}
(The plus sign choice in (\ref{P4.15}) would correspond to space like geodesics outside the horizon)
\begin{equation}\label{P4.16}
and~~~~~~~~~\dot{v}=\frac{1}{A+r^4 \dot{r}}.
\end{equation}
We can see that the value of $A$ is less than zero for the space like geodesic. Then $\dot{r}$ and $\dot{v}$ are both less than zero and there exist only positive terms in (\ref{P4.9}). From the equation (\ref{P4.15}), we get
\begin{equation}\label{P4.17}
\frac{V_{\Sigma}}{4 \pi}=\lambda_f=\int^{2ma(t)}_0 dr \frac{r^4}{\sqrt{A^2+r^4 f(r,t)}}.
\end{equation}
Equation (\ref{P4.17}) indicates some restriction on $A$ as
\begin{equation}\label{P4.18}
A^2+r^4f(r,t)>0\rightarrow A^2>-r^4_V f(r_V,t)=\frac{27}{16}m^4a^4(t)=A_a^2.
\end{equation} 
By analysing the polynomial, we get the last condition. It gets a maximum value at $r_V=\frac{3}{2}m a(t)$. Also $r=0$ and $r=2ma(t)$ are the roots. Here, it is positive in this said range.

If the radial value is constant, the equation (\ref{P4.15}) and (\ref{P4.16}) can be written as
\begin{equation}\label{P4.19}
A^2=-r^4 f(r,t)
\end{equation}
and
\begin{equation}\label{P4.20}
\dot{v}=\frac{1}{A}.
\end{equation}
For every constant value of $r$, we have a solution in the range $0<r<2ma(t)$, since $-r^4f(r,t)>0$. Surfaces for the constant value of $r$ are identically stationary (maximal) points of the volume ($V_{\Sigma}[\gamma]$). By integrating the equation (\ref{P4.20}), we get 
\begin{equation}\label{P4.21}
\lambda_f=A(v_f-v).
\end{equation}
Between two given $v$ in the $r=constant$ surface, we get the largest volume when $A$ is large ,i.e., we get $A=A_a$ for $r=r_V$. These assumptions give the basic derivation of the asymptotic volume. The derivation is done in the rest part of this letter.

Now, we are constructing the volume for large $v$. ${\cal{S}}$ is located at the point $(v,2m a(t))$ up to $(v_f,0)$. In that range, the proper length of the space like geodesic should increase monotonically. The $v$ coordinates can be simply estimated, where $\Sigma_v$ reaches $r=0$: it must be onwards the construction of the singularity, whereas the accessible volume increases for this, and we can choose $v_f=0$ except any significative error for the large $v$ limit. So, we choose the end point of $\gamma$ at the coordinate $(0,0)$. 
 
Can we estimate the path for which the volume is maximized? The final inspection is that for maximization of $\lambda_f$, when $v$ is very large, the geodesic must allocate the maximum probable time for the radius $r$. Also the line element is longer and occurs to have a maximum. So we may relate the geodesic with a final and an initial transients and an intermediate long phase (where $\dot{r}\sim 0$). Hence the auxilliary line element (\ref{P4.11}) is given by
\begin{equation}\label{P4.22}
ds_{{\cal{M}}_{aux}}\sim -\sqrt{-r^4 f(r,t)}dv,
\end{equation}
where increment of $v$ is improving the approximation. The minus sign is taken because $dv<0$. To get the maximum length, we have to maximise $ds/dv$, where the steady phase of the geodesic runs for the value of $r$. Then we get 
\begin{equation}\label{P4.23}
\frac{d}{dr}\sqrt{-r^4 f(r,t)}=0.
\end{equation}
The solution of the above equation gives the maximum value of the polynomial $-r^4f(r,t)$, which already is mentioned as $r_V$
\begin{equation}\label{P4.24}
r_V = \frac{3}{2}m a(t).
\end{equation}
So for large $v$, we get the maximised spherically symmetric spacelike surface which is constructed by a long stretch at near to the constant radius $r_V=\frac{3}{2}ma(t)$, joined from $r=0$ on one side to the horizon given by $r=2ma(t)$ to the opposite extremity by transients.
Hence, 
\begin{equation}\label{P4.25}
V\approx -4\pi r_V^4 f(r_V,t)v=3\sqrt{3} \pi m^2 a^2(t)v,
\end{equation}
it is the composite form of the equation (\ref{P4.13}) and (\ref{P4.21}) for $v_f=0$.
\begin{figure}[ht]
\begin{center}

~~~~~~~~~~~~~~~~~~~~~~~~~~~~Fig.1(i)~~~~~~~~~~~~~~~~~~~~~~~~~~~~~~~~~~~~~~~~~\\
\vspace{0.5cm}
\includegraphics[height=2.5in, width=3.2in]{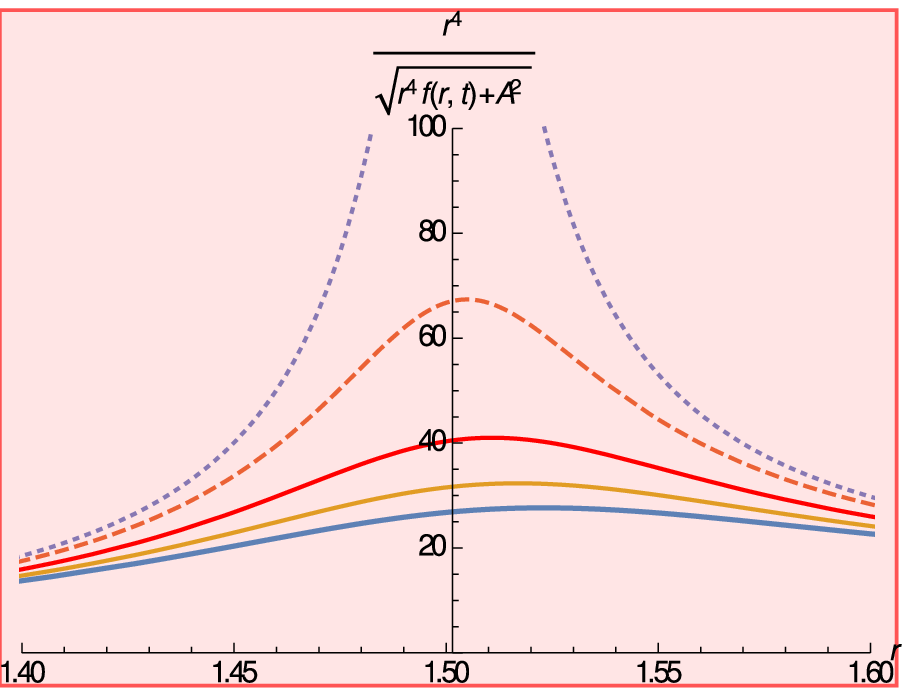}\\
\vspace{0.1cm}
Fig 1(i) : $a(t)=(1+a_1 t),~A \rightarrow A_a$, where $a_1=0.01,~m=1,$ $t=1$, $r_V=1.5015$ and origin is at $(1.5015,0)$
\end{center}
\end{figure}
\newpage
\begin{figure}[ht]
\begin{center}

~~~~~~~~~~~~~~~~~~~~~~~~~~~~Fig.1(ii)~~~~~~~~~~~~~~~~~~~~~~~~~~~~~~~~~~~~~~~~~\\
\vspace{0.5cm}
\includegraphics[height=2.5in, width=3.2in]{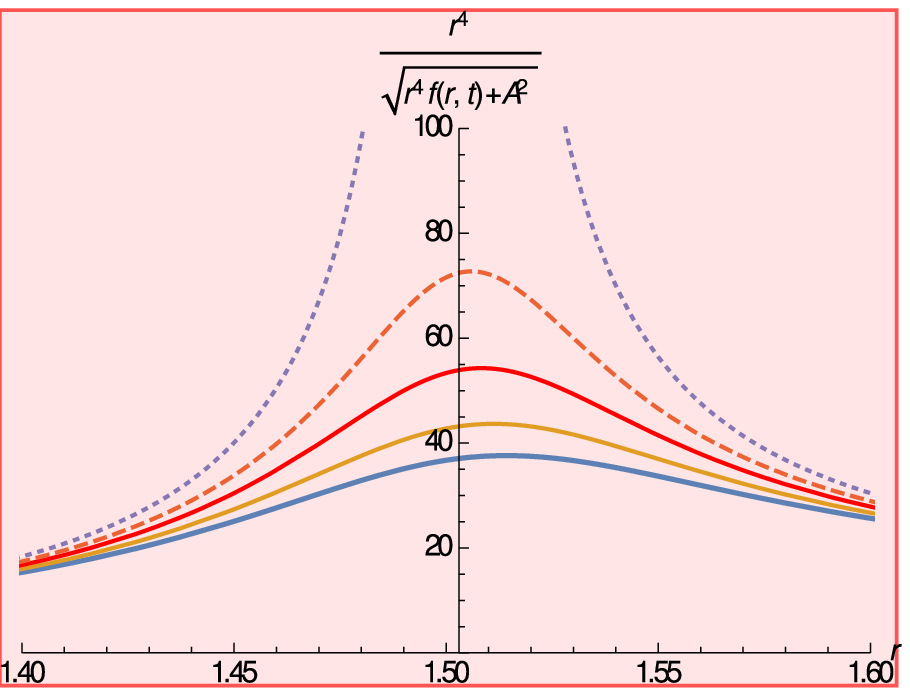}\\
\vspace{0.1cm}
Fig 1(ii) : $a(t)=(1+a_1 t+a_2 t^2),~A \rightarrow A_a$, where $a_1=0.001,~a_2=0.001,~m=1$, $t=1$, $r_V=1.503$ and origin is at $(1.503,0)$
\end{center}
\end{figure}

\begin{figure}[ht]
\begin{center}

~~~~~~~~~~~~~~~~~~~~~~~~~~~~Fig.1(iii)~~~~~~~~~~~~~~~~~~~~~~~~~~~~~~~~~~~~~~~~~\\
\vspace{0.5cm}
\includegraphics[height=2.5in, width=3.2in]{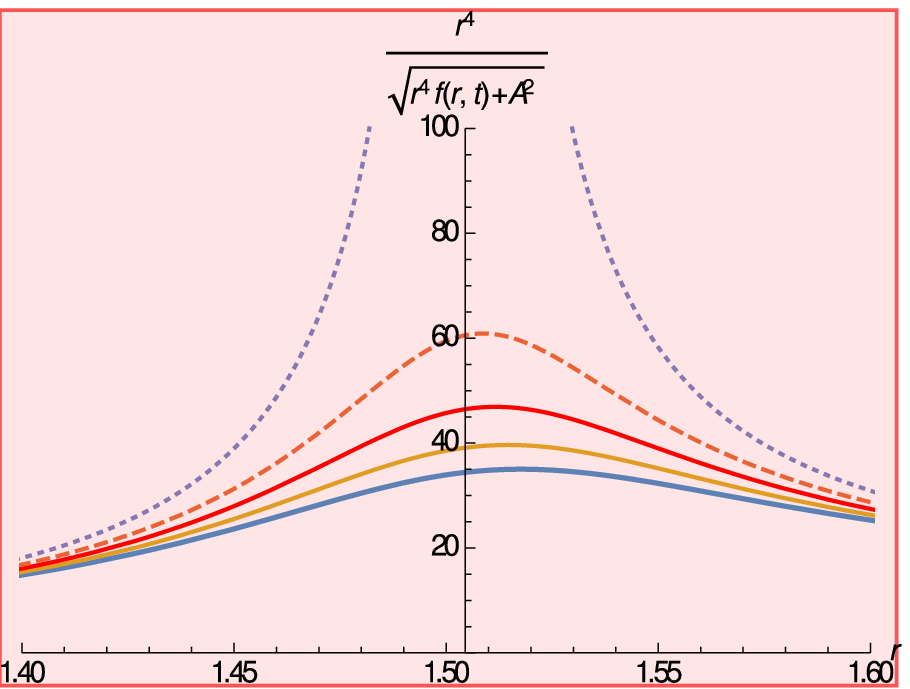}\\
\vspace{0.1cm}
Fig 1(iii) : $a(t)=(1+a_1 t+a_2 t^2+a_3 t^3),~A \rightarrow A_a$, where $a_1=0.001,~a_2=0.001=a_2=0.001,~m=1$, $t=1$, $r_V=1.5045$ and origin is at $(1.5045,0)$
\end{center}
\end{figure}
From the above three figures we are trying to show that if we are considering $A$ to tend to $A_a$ then the total volume increases. The surface $\Sigma$ has some regions depending on the radius of the black hole. It has two transient regions for the lower and larger value of the radius. But when the radius is equal to $\frac{3}{2}m a(t)$, it has a long steady region. So from a small region neighboring $r_V=\frac{3}{2}m a(t)$, we get a major contribution on the volume increasing as $A \rightarrow A_a$.

Here, we get the volume which is modified form of CR volume which might exist inside the black hole. It is necessary to determine the number of modes of quantum field which can be combined in this volume. For the standard quantum statistical method\cite{Cowan1}, one can determine the number of quantum states of some volume for one specific phase space which are labeled by $\lambda,\theta,\phi,p_{\lambda},p_{\theta} ~and~p_{\phi}$. The uncertainty relation of quantum mechanics is stated as  $\bigtriangleup x_i \bigtriangleup p_i\sim 2\pi$, one quantum state be similar to a `cell' of volume equal to $(2\pi)^3$ in the phase space. Then the number of states is given by
\begin{equation}\label{P4.26}
\frac{d\lambda d\theta d\phi dp_{\lambda} dp_{\theta} dp_{\phi}}{(2\pi)^3}~~.
\end{equation}
We have to calculate the integral of density of states in the phase space. Therefore we assume the massless scalar field $\Phi$ in the space-time with the coordinates\cite{Zhang1} 
\begin{equation}\label{P4.27}
ds^2= -dT^2 + (-f(r,t)\dot{v}^2 + 2 \dot{v} \dot{r})d\lambda^2 + r^2d\theta^2 + r^2 sin^2\theta d\phi^2~~,
\end{equation}
This metric is equivalent to the form 
\begin{equation}\label{P4.28}
ds^2= -dT^2 + H(T) d\lambda^2 + r(T)^2d\theta^2 + r(T)^2 sin^2\theta d\phi^2
\end{equation}
which implies that the hypersurface is dynamical for the defined time $T$ in the interior of the BH. The interior of the BH sometimes interpreted as cosmological model for this reason\cite{Carroll1} and it evolves to the singularity of BH. Due to the physical quantities, statistical mechanies can not be defined clearly in the dynamical background.  We must take a static background to gain the statistical property of the scalar field in the interior of the BH. According to \cite{Christodoulou1} the hypersurface at constant $r$ can be identified as static which $v>>m$ and $\dot{r}=0$. So the hypersurface is maximised as $r=r_{max}=\frac{3}{2}m a(t)$ where $t$ is large enough\cite{Estabrook1}. Near the maximal hypersurface, the proper time between neighbouring hypersurface tends to $0$. As $t$ increases, there is no evolution found. Thus our calculation is free from effect of the non-static character of the metric, since it is determined on approximately $T=constant$. So we will use usual method in the curved spacetime to know the motion of the scalar field in the interior of BH.

The scalar field $\Phi$ can be written as 
\begin{equation}\label{P4.29}
\Phi=exp\{-iET\} exp\{iI(\lambda, \theta, \phi)\}
\end{equation}  
by using WKB approximation.

Substituting this into the Klein-Gordon equation in curve spacetime,
$$\frac{1}{\sqrt{-g}} \partial_{\mu}(\sqrt{-g}g^{\mu \nu}\partial_{\nu}\Phi)=0,$$
we get
\begin{equation}\label{P4.30}
E^2 - \frac{1}{-f(r,t)\dot{v}^2 + 2\dot{v} \dot{r}}p^2_{\lambda} - \frac{1}{r^2}p^2_{\theta} - \frac{1}{r^2 sin^2\theta}p^2_{\phi}=0,
\end{equation}
where $p_\lambda = \frac{\partial I}{\partial \lambda},~p_\theta = \frac{\partial I}{\partial \theta},~p_\phi = \frac{\partial I}{\partial \phi}$. From the equation (\ref{P4.26}), the number of states with energy ($<E$) is obtained as
$$g(E)= \frac{1}{(2\pi)^3}\int d\lambda d\theta d\phi dp_{\lambda} dp_{\theta} dp_{\phi}= \frac{1}{(2\pi)^3}\int \sqrt{-f(r,t)\dot{v}^2+2\dot{v}\dot{r}} d\lambda d\theta d\phi  \int \sqrt{E^2-\frac{1}{r^2}p^2_{\theta}-\frac{1}{r^2 sin^2\theta}p^2_{\phi}} dp_{\theta} dp_{\phi}$$
\begin{equation}\label{P4.31}
= \frac{1}{(2\pi)^3}\int \sqrt{-f(r,t)\dot{v}^2+2\dot{v}\dot{r}} d\lambda d\theta d\phi \left(\frac{2\pi}{3}E^3r^2sin\theta\right)=\frac{E^3}{12{\pi}^2}\left[4\pi\int d\lambda \sqrt{r^4\left(-f(r,t)\dot{v}^2+2\dot{v}\dot{r}\right)}\right]= \frac{\sqrt{3} E^3}{4 \pi}m^2 a^2(t)v
\end{equation}
Thus the number of quantum states is proportional to the volume which we have obtained as $V$. Ignoring the exotic feature of $V$ we can continue to compute the free energy at inverse temparature $\beta$,
\begin{equation}\label{P4.32}
{\cal{F}}(\beta) = \frac{1}{\beta} \int dg(E)~ln(1-e^{-\beta E}) = - \int \frac{g(E)dE}{e^{\beta E}-1}= - \frac{V}{12 \pi^2} \int \frac{E^2 dE}{e^{\beta E}-1} = - \frac{\pi^3 }{20\sqrt{3} \beta^4}m^2 a^2(t)v.
\end{equation}
Also, the entropy is 
\begin{equation}\label{P4.33}
S = \beta^2 \frac{\partial {\cal{F}}}{\partial \beta} = \frac{\pi^3 }{5\sqrt{3} \beta^3}m^2 a^2(t)v.
\end{equation}
This entropy is calculated normally but it is inevitable considering Hawking radiation into our calculation. For consideration of Hawking radiation we need the mass loss rate from the black hole which we get from Stefan-Boltzmann law for Schwarzschild BH,
\begin{equation}\label{P4.34}
\frac{dm}{dv} = - \frac{1}{\gamma m^2}, ~\gamma>0,
\end{equation}
where $\gamma$ is constant which has no effect directly in the calculation. From this law we have for a black hole with mass $m$, 
\begin{equation}\label{P4.35}
v \sim \gamma m^3.
\end{equation}
It satisfies our requirment that is $v>>m$. Therefore, we get the entropy S as
\begin{equation}\label{P4.36}
S \sim \frac{\gamma m^2 a^2(t)}{(5\sqrt{3} \times 8^3)} = \frac{\gamma}{(10\sqrt{3} \times 8^4)\pi}{\cal{A}}a^2(t),
\end{equation}
where we are considering the inverse temperature for a Schwarzchild black hole, $\beta = T^{-1} = 8 \pi m$ and ${\cal{A}} = 16 \pi m^2$ is the surface area of Schwarzschild black hole. This is an intriguing and surprising result that the entropy is proportional to the square of $a(t)$ and the surface area of the BH horizon that covers the volume. Also this result depends on the relation (\ref{P4.34}), which was demonstrated in\cite{Massar1} to hold so long as the mass of the Schwarzchild BH is greater than the Planck mass.

We can compare our metric given by eqn (\ref{P4.1}) with a standard non commutative black hole \cite{32} as below
$$\gamma\left(\frac{3}{2}, \frac{r^2t}{4\theta}\right)\\
= \frac{r^6 t^3}{192 \theta^3} \left[ 1- \frac{r^2 t}{2.66 \theta} + \frac{r^4 t^2}{53.33 \theta^2} - \frac{r^6 t^3}{768 \theta^3} +... + \frac{r^2n t^n}{n! 4^n n^2 \theta^n}\right]
= \frac{r^6 t^3}{192 \theta^3} \left[ 1+a_1t+a_2t^2+...+a_nt^n \right],$$
where $a_1= \frac{r^2 }{2.66 \theta}$,$ a_2= \frac{r^4}{53.33 \theta^2}$,..., $a_n=\frac{r^2n }{n! 4^n n^2 \theta^n}$. Then we write $f(r,t)$ of eqn (\ref{P4.1}) as
$$ f(r,t)= 1-\frac{768 m \theta^3}{r^7t^3 \sqrt{\pi}} \gamma\left(\frac{3}{2}, \frac{r^2t}{4\theta}\right),$$
for a large black hole if we assume $\theta= r^2 t$ we get back a static symmetric trivial black hole solution.

The entropy associated to the non commutative volume remains insufficient for a statistical interpretation of the black hole entropy if $v$ only is accounted for the first evaporation stage. To incorporate the second evaporation stage we will follow the reference \cite{Nicolini}. This work deals with noncommutative space time in the point of view of final explosion/evaporation stage with a diverging temperature (when $M_f\rightarrow M_0$). In this regime the approximation of the temperature can be taken as $T_h\simeq \alpha (M_f-M_0)$ with $\alpha=\left.\frac{dT_h}{dM}\right|_{r_h=r_0}$. Following the analysis of \cite{32} we have the expression for large $V$ as
$$v\sim \frac{1}{(M_f-M_0)^3}.$$
It is quite clear that the final evaporation stage will require an infinite time. This phenomenon will be consisted with the third law of thermodynamics, the statement of which depicts that the zero temperature state can not be reached with a countable number of steps or within a finite time. Following the references \cite{1,2,3,4} we find 
$$V_{NCR}\simeq \frac{3\sqrt{3} \pi M_f^2\left(1-\frac{2r_n}{\sqrt{\pi \theta}}\exp \left\{-\frac{r_n^2}{4\theta}\right\}\right)}{\left(M_f-M_0\right)^3}$$
for $v\rightarrow \infty$, when $M_f \rightarrow M_0$, the $V_{NCR}$ is divergent. This is a black hole with an infinite volume wrapped by a finite horizon.

So we obtain from $S_{NCR}=\frac{\pi^2V_{NCR}}{45\beta_h^3}$ the expression of $S_{NCR}$ as $$S_{NCR}\sim \eta \alpha^3A_h$$
with $$\eta=\frac{\sqrt{3}\pi^2\left(1-\frac{2r_n}{\sqrt{\pi \theta}}\exp \left\{-\frac{r_n^2}{4\theta}\right\}\right)}{15\times 16 \left(1-\frac{2M}{\sqrt{\pi \theta}}\exp \left\{-\frac{M^2}{\theta}\right\}\right)^2}\simeq 0.05$$
when $\alpha \rightarrow \infty$ when $r_h\rightarrow r_0$, from $\left.\frac{dM}{dr_h}\right|_{r_h=r_0}=0$ we see this expression approaches infinity as well.

So We can conclude that the non commutative black hole can possess an infinite CR volume. This is a memeber of a class of black holes with finite surface, i.e., entropy, but an infinite interior. This conclusion supports the fact that the black hole entropy may be independent of the interior of a black hole.

We have observed that for volume inside a sphere which is a spherically symmetric context is the maximal volume for a spacelike spherically symmetric $3d$ hyper surface covered by the sphere. Also we have calculated this proper volume for a black hole which is spherically symmetric created by a collapsed object. Also we have construct that the volume inside the black hole is defined by the equation (\ref{P4.25}). The attractive view of this result is that $V$ is large and it is proportional to the function $a(t)$ and $v$, i.e., it is increases with a time function. But Christodoulou and Rovelli\cite{Christodoulou1} have found that $V$ increases linearly with time since the black hole collapse.

Also we have computed the entropy in this volume with a standard statistcal method. The entropy is proportional to the square of $a(t)$ and the entropy is increasing with time.

\vspace{.2 in}
{\bf Acknowledgement:}
This research is supported by the project grant of Goverment of West Bengal, Department of Higher Education, Science and Technology and Biotechnology (File no:- $ST/P/S\&T/16G-19/2017$). SD thanks Goverment of West Bengal, Department of Higher Education, Science and Technology and Biotechnology for Non-NET Fellowship. RB thanks Inter University Center for Astronomy and Astrophysics(IUCAA), Pune, India for Visiting Associateship. PC thanks CSIR, INDIA for awarding JRF.

\end{document}